\documentclass[psfig]{llncs}
\usepackage{multicol} % needed for the list of participants
\usepackage{makeidx}  % allows for indexgeneration
\usepackage{psfig}
\usepackage{amssymb}
\begin{document}                

\title{Metastable States in CA models for Traffic Flow}

\author{R. Barlovic\inst{1}, L. Santen\inst{2},
  A. Schadschneider\inst{2}, M. Schreckenberg\inst{1}}

\institute{Theoretische Physik FB10\\
  Gerhard-Mercator Universit\"at Duisburg \\
  D-47048 Duisburg, Germany \\
  $ $ 
  \and Institut f\"ur Theoretische Physik\\ 
  Universit\"at zu K\"oln \\ D-50937 K\"oln, Germany}

\maketitle           
\begin{abstract}
Measurements of traffic flow show the existence of metastable states of
very high throughput. These observations cannot be reproduced by the
CA model of Nagel and Schreckenberg  (NaSch model), not even 
qualitatively.  Here we present two variants on the NaSch model with
modified acceleration rules ('slow-to-start' rules). Although these
models are still
discrete in time and space, different types of metastable states
can be observed. 
\end{abstract}

\section{Introduction}

Simulation results of the CA model for traffic flow introduced by Nagel and
Schreckenberg \cite{NaSch} (NaSch model, for a  detailed explanation
see \cite{NaSch,duiproc}) show a reasonable agreement
with experimental 
data, although the model definition includes only a few simple
update rules. Due to its simplicity it can be used very efficiently for 
computer simulations, but also an analytical treatment is 
possible \cite{duiproc}. 

One important experimental result \cite{Kerner} which cannot be gained
from the NaSch model is the occurance of metastable states near the
density of maximum flow. 
Very recently such states have been found in a modified version
of the NaSch model where a limited braking capacity of the cars
and a continuous space is considered \cite{kraussmod}. The continuous space
allows very small amplitudes of the fluctuations compared to the 
discrete NaSch model. Therefore one could ask if a continuous space is
neccessary to obtain metastable states.
 
Here we present the results of a numerical investigation of two
variants on the NaSch model with modified acceleration rules, which
are still discrete in time and 
space. First, we assigned a larger braking probability to stopped cars
and second, we analysed the modification suggested by Takayasu and
Takayasu \cite{T2mod}. It turns out that metastable  states can be
found in both cases.
 
\section{NaSch model with 'slow-to-start' rule}

In this part of the investigation we consider the NaSch model with
maximum velocity $v_{max}=5$ and braking probability $p=0.01$ of the moving
cars. In contrast to the original model, a higher value
$p_s = 0.5$ of the braking probability is assigned to stopped cars.
  
We performed simulation runs for periodic systems containing $L=10000$
lattice sites. This system size is sufficient to exclude
finite size effects for the chosen set of parameters.     
\begin{figure}[h]
  \centerline{\psfig{figure=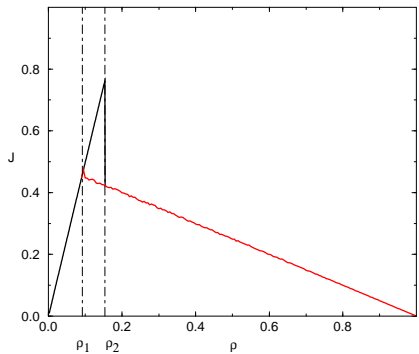,bbllx=20pt,bblly=20pt,bburx=550pt,bbury=450pt,height=5cm}}
\caption{\protect{Fundamental diagram of the modified NaSch model 
($v_{max}=5$, $p_{s}=0.5$, $p =0.01$, $L=10000$).}}
\label{fund_nasch}
\end{figure} 
Fig. \ref{fund_nasch} shows the fundamental diagram of the modified
model. Obviously the average flow $J(\rho)$ can take two values in the
density interval between $\rho_{1}$ and $\rho_{2}$ dependent on the
chosen initialisation. The larger values of the average flow can be
obtained using a homogeneous initialisation of the system. The lower
branch is obtained starting from a complete jammed state. Moreover
varying the particle number adiabatically, one can trace a hysteresis
loop. One gets the upper branch by adding cars to the stationary state
with $\rho < \rho_1$ and the lower one by removing cars from the
stationary state with $\rho > \rho_2$.
 For a fixed value of $p$, $\Delta J =
J(\rho_{2})-J(\rho_{1})$ depends linearly on $p_s$ for wide a range of
parameters. 

The microscopic structure of the jammed states in the modified model
differs from those found in the NaSch model. While jammed states in
the NaSch model contain with an exponential size-distribution
\cite{duiproc},
one can find phase separation in the modified model. The reason for
this behaviour is the reduction of the outflow from a jam. Therefore
the density in the free flow regime is smaller than the density of
maximum flow and cars can propagate freely in the low density part of
the lattice. Due to the reduction of the density in the free flow
regime, no spontaneous formation of jams is observable in the stationary
state, if fluctuations in the free flow regime are rare. 

This picture can be confirmed by a simple phenomenological
approach. Obviously the flow in the homogeneous branch is given by $J_h
= \rho (v_{max}-p) = \rho v_{f}$, because every car can move with its
desired velocity $v_{f}$. Assuming that the high density states
are phase separated, we can obtain the second branch of the fundamental
diagram. The phase separated states consist of a large jam and a free
flow regime, where each car moves with velocity $v_f$. The density in
the free flow regime $\rho_f$ is determined by the average waiting
time $T_w = \frac{1}{1-p_s}$  of the
first car in the jam  and $v_f$, because
neglecting interactions between cars the average distance of two
consecutive cars is given by $\Delta x = T_w v_{f} + 1 = \rho_f^{-1}$. Using
the normalisation $L = N_J + N_F \Delta x$ ($N_{F(J)}$ number of cars
in the free flow regime (jam)) we find that the flow is given by 
\begin{equation}
  \label{jjam}
  J_s(\rho) = (1-\rho)(1-p_s).
\end{equation}
Obviously $\rho_f$ is precisely the branching density $\rho_1$, because
for densities below $\rho_f$ the jamlength is zero. 
It should be noted that this approach is only valid if ($p_s \gg p$) and
$v_{max} >1$ holds. Small values of $p$ have to be considered in order
to avoid interaction between cars due to the velocity fluctuations and
$v_{max} > 1$, because otherwise cars can stop spontaneous in the free
flow regime and therefore initiate a jam.

\begin{figure}[h]
 \centerline{\psfig{figure=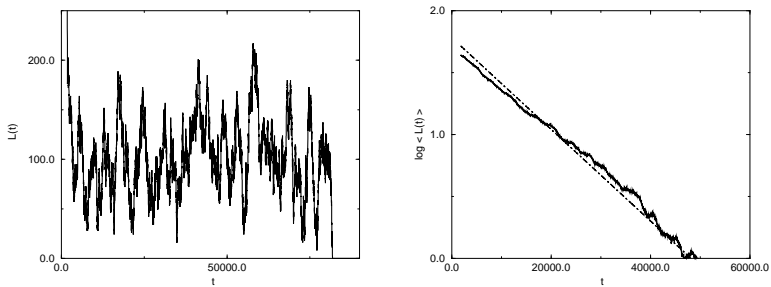,bbllx=0pt,bblly=0pt,bburx=1170pt,bbury=530pt,height=5cm}}
 \caption{\protect{Time-dependence of the jamlength for $\rho
     =0.095$. The left part of the figure shows the time evolution of
     the length L(t) of one sample. The average over 10000 samples
     (right part of the figure) 
     shows that the jamlength decays exponentially.}}
\label{jam_length}
\end{figure}

Measurements of the average flow show that the lower branch of the
fundamental diagram is not stable near the density $\rho_1$. Therefore
we performed a more detailed stability analysis of the homogeneous and
the jammed state near  $\rho_1$ and  $\rho_2$.  Near $\rho_1$ the
large jam present in the initial configuration 
resolves, the average length decays exponentially
in time (Fig. \ref{jam_length}). It should be noted that this behaviour is not
the consequence of a continuous "melting" of the large jam. In
contrast, the jamlength is strongly fluctuating without any systematic
time-dependence.  Once a homogeneous
state without a jammed car is reached, no jam appears again. Therefore
the homogeneous state is stable near $\rho_1$. 
\begin{figure}[h]
  \centerline{\psfig{figure=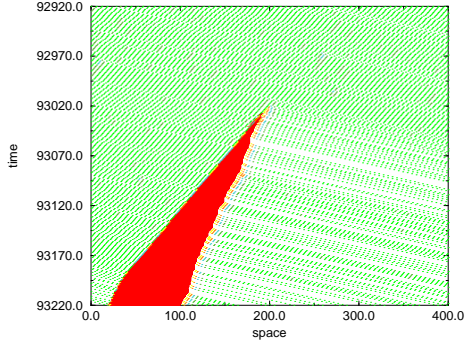,bbllx=0pt,bblly=0pt,bburx=570pt,bbury=530pt,height=6cm}}
  \caption{\protect{Space-time diagram of the modified NaSch model for
    $\rho=0.15, L=400, p =0.01 $ and $p_s=0.5$. The homogeneous initial
      state is not destroyed 
      immediately, but after approximately 90000 lattice updates. In
      the outflow regime of the jam the density is reduced compared to
      the average density.}}
\label{st_diag}
\end{figure}

Analogous to the metastable jammed states near $\rho_1$, homogeneous
initialisations for densities slightly above $\rho_2$  lead to
metastable homogeneous states with short lifetimes. Fig. \ref{st_diag}
shows the spontaneous formation of jams due to velocity fluctuations. The
finite lifetimes of the homogeneous states are the qualitative
difference between this model and the cruise-control limit
\cite{Pac}, where the
time evolution of homogeneous states at low densities is completely
deterministic.

\section{T$^{2}$ model}

Takayasu and Takayasu (T$^{2}$) \cite{T2mod} suggested a cellular
automaton model with another type of a slow-to-start rule, which is defined as
follows: A standing car with exactly one empty cell in front of it
accelerates with probability $q_t = 1-p_t$, while all other cars
accelerate deterministicly. The other update rules of the NaSch model
are unchanged.  Due to this modification already for $v_{max}=1$ the
particle-hole symmetry is  broken.    
\begin{figure}[h]
 \centerline{\psfig{figure=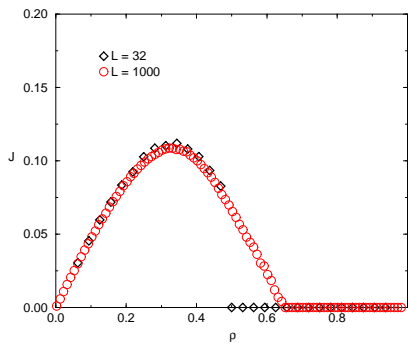,bbllx=0pt,bblly=0pt,bburx=550pt,bbury=450pt,height=5cm}}
\caption{\protect{Fundamental diagram of the $T^2$ model ($v_{max}= 1$,
   $p = 0.5$, $p_t =1$) for two system sizes. For densities slightly
  above $\rho = 0.5$ the stationary state could only be reached for
  the small system.}}
\label{T2_fund}
\end{figure}
In Fig. \ref{T2_fund}  we show the fundamental diagram for
$v_{max}=1$, $p=0.5$ and $p_t =1$, i.e. stopped cars can only
move if there are at least two empty cells in front of it.  Obviously
completely blocked states exist for densities $\rho \geq 0.5$, where
the number of empty cells in front of each car is smaller than
two. Since fluctuations are absent in those states, they
have an infinite lifetime. Therefore the flow in the stationary state is zero.

Although the states with a finite flow are not stationary, typically one has to
 perform an extremely large number of update steps until the flow
 vanishes for large system 
 sizes and densities  slightly above $\rho = 0.5$, because the number
 of blocked configurations is very 
 small compared to the total number of configurations. Precisely at
 $\rho = 0.5$, the blocked state is unique and the typical time to
 reach this state diverges exponentially with the system size. Therefore
 we used very small systems in order to obtain the lower branch of
 the fundamental diagram. 
 
 Another interesting feature is the form of the fundamental
 diagram. There is  some experimental evidence that in certain
 situations the shape of the fundamental diagram differs from the
 standard convex form. This behaviour
 of the average flow can be easily obtained tuning  the parameter
 $p_t$ \cite{duiproc,slow_comf}.  

\section{Summary}

The numerical analysis of two modifications of the NaSch model shows
that metastable states can also be found in CA models for
traffic flow.

The NaSch model with 'slow-to-start' rule shows the coexistence of 
phase separated and homogeneous states in a density interval near the
density $\rho_2$ of maximum flow. Near $\rho_2$ interactions between
cars become important and one can find spontaneous formation of
jams. Therefore the reduction of the density in the outflow regime
of a jam leads to stable phase separated states. The reduction of interactions
between cars in the free flow regime can be confirmed by a
phenomenological approach, which gives very accurate results. In
contrast to the cruise-control limit of the NaSch model \cite{Pac}, where also
metastable states can be found, fluctuations are present in both
coexisting states. Finally we mention that this
modified version of the NaSch model can reproduce the fundamental
diagram of the spacecontinuous approach by Krau{\ss}
et. al. \cite{kraussmod}. 
   
In the $p_t =1$ limit of the $T^2$ model we find metastable states for
densities beyond $\rho =0.5$. For $\rho \gtrsim 0.5$ states with a
finite flow have very large lifetimes, but if the system reaches a
blocked state these states are stable because of the absence of
fluctuations. Therefore the $p_t =1$ limit of the $T^2$ model is
complementary to the cruise-control limit of the NaSch model, where
one finds the absence of fluctuations for homogeneous states at low
densities. 

These modifications of the NaSch model show that CA models for traffic can
reproduce a broad spectrum of experimental results although the update
rules oversimplify the individual behaviour of the drivers. 
%
% ---- Bibliography ----
%

\end{document}